\documentclass{article}
\usepackage{psfig}
\begin{document}
\title{Time delay as a key to Apoptosis Induction
       in the p53 Network}
\author{G. Tiana$^{1,2}$, M. H. Jensen$^2$ and K. Sneppen$^3$\\
        $^1$Department of Physics, Univeristy of Milano,\\
        via Celoria 16, 20133 Milano, Italy\\
        $^2$The Niels Bohr Institute \\
        Bledgamsvej 17, 2100 Copenhagen, Denmark\\
        $^3$  Department of Physics, Norwegian University of Science
and Technology,\\ N-7491, Trondheim, Norway.  } 
\maketitle

\begin{abstract}
A feedback mechanism that involves the proteins p53 and mdm2, 
induces cell death as a controled response to severe 
DNA damage. A minimal model for this mechanism demonstrates that the 
respone may be dynamic and connected
with the time needed to translate the mdm2 protein.
The response takes place if the dissociation
constant $k$ between p53 and mdm2 varies from its normal value. 
Although it is widely believed
that it is an increase in $k$ 
that triggers the response, we show
that the experimental behaviour is better described by a decrease in
the dissociation constant.  
The response is quite robust upon changes in the parameters of the
system, as required
by any control mechanism, except for few weak points, which 
could be connected with the onset of cancer.
\end{abstract}

\bigskip
\centerline{PACS: 87.16.Yc}

\newpage
\section{Introduction}
In healthy
cells, a loopback mechanism involving the protein p53
is believed to cause growth arrest and apoptosis
as a response to DNA damage \cite{vogelstein,shair,agarwal,baror}.
Mutations in the sequence of p53 that potentially interfere
with this mechanism
have been observed to  
lead to the upraise of cancer \cite{greenblatt,gottlieb}.

Under normal conditions the amount of p53 protein in the cell 
is kept low by a genetic network built of
the mdm2 gene, the mdm2 protein and the p53 protein itself. p53 is
produced at 
a essentially 
constant rate and promotes the expression of the mdm2 gene \cite{haupt}.
On the other hand, 
the mdm2 protein binds to p53 and promotes its degradation
\cite{kubbutat}, decreasing its
concentration. When DNA is damaged, a cascade of events causes
phosphorylation 
of several serines in the p53 protein, which modifies its binding
properties 
to mdm2 \cite{unger}. As a consequence, the cell experiences a sudden
increase 
in the concentration of p53, which activates a group of genes (e.g.,
p21, 
bax \cite{eldeiry}) responsible for cell cycle arrest and apoptosis.
This increase 
in p53 can reach values of the order of 16 times the basal
concentration \cite{oliner}.

A qualitative study of the time dependence of the concentration of p53 
and mdm2 has been carried out in ref. \cite{haupt}. Approximately one
hour 
after the stress event (i.e., the DNA damage which causes
phosphorylation of p53 serines), a peak in the concentration of p53 is 
observed, lasting for about one hour.
This peak partially overlaps with the peak in the concentration of 
mdm2, lasting from $\approx 1.5$ to $\approx 2.5$ hours after the 
stress event. Another small peak in the concentration of p53 is
observed after several hours.

The purpose of the present work is to provide the simpest mathematical
model 
which describes all the known aspect of the p53--mdm2 loop, and to
investigate 
how the loop is robust to small variations to the ingredients of the
model. 
The "weak points" displayed by the system, namely those
variations in some parameters which cause abrupt changes in the overall
behaviour of the
loop, are worth to be investigated experimentally because 
they can contain informations about how a cell bacomes tumoral.

The model we suggest is described in Fig. 1. 
The total number of p53 molecules, produced at constant rate $S$, is
indicated with $p$.
The amount of the complexes built of p53 bound to mdm2 is called $pm$. 
These complexes cause the degradation 
of p53 (through the ubiquitin pathway), at a rate $a$, while mdm2
re--enters 
the loop. Furthermore, p53 has a spontaneous decay rate $b$.
The total number of mdm2 proteins is indicated as $m$.
Since p53 activates the expression of the mdm2 gene, the
production rate of mdm2 is proportional (with constant $c$) to the
probability that 
the complex p53/mdm--gene is built.
We assume that the complex p53/mdm2--gene is at equilibrium with its
components, where $k_g$ is
the dissociation constant and only free p53 molecules (whose amount is
$p-pm$) can
participate into the complex. The protein mdm2 has a decay
rate $d$. The constants
$b$ and $d$ describe not only the spontaneous degradation of the
proteins, 
but also their binding to some other part of the cell, not described 
explicitely by the model.
The free proteins p53 and mdm2 are considered to be at equilibrium with
their bound complex pm, 
and the equilibrium constant is called $k$. 

The dynamics of the system can be described by the equations
\begin{eqnarray}
\label{eq1}
\frac{\partial p}{\partial t} & = & S - a\cdot pm - b\cdot p
\\\nonumber
\frac{\partial m}{\partial t} & = & 
c\frac{p(t-\tau)-pm(t-\tau)}{k_g+p(t-\tau)-pm(t-\tau)} - d\cdot m
\\\nonumber
pm & = & \frac{1}{2}\left((p+m+k)-\sqrt{(p+m+k)^2-4p\cdot m}\right). 
\end{eqnarray} 
In the second equation we allow a delay $\tau$ in the production of
mdm2, due to 
the fact that the transcription and translation of mdm2 lasts for some 
time after that p53 has bound to the gene.

The choice of the numeric parameters is somewhat difficult, due to the
lack of reliable
experimental data. The degradation rate through ubiquitin pathway has
been estimated to be
$a\approx 3\cdot 10^{-2}s^{-1}$ \cite{wilkinson}, while the spontaneous
degradation of p53 is 
$\approx 10^{-4} s^{-1}$ \cite{haupt}. The dissociation constant between
p53 and mdm2 is $k\approx 180$ \cite{kussie} (expressed as number of
molecules, assuming for 
the nucleus a volume of $0.6 \mu m^3$), and
the dissociation constant between p53 and the mdm2 gene is $k_g\approx
28$ \cite{bala}.
In lack of detailed values for the protein production rates, we have
used typical 
values, namely $S=1 s^{-1}$ and $c=1 s^{-1}$. The degradation rate of
mdm2 protein has been chosen of the order of $d=10^{-2} s^{-1}$ to keep
the stationary amount of mdm2 of the order of $10^2$. 

The behaviour of the above model is independent on the volume in which
we assume the reaction takes place. 
That is, multiplying $S$, $c$, $k_g$ and $k$ by the same constant
$\omega$ gives exactly
the same dynamics of the rescaled quantities $\omega p$ and $\omega m$.
Futhermore, due to the fact that the chosen parameters put the
system in the saturated regime, an increase in the producing rates
$S$ and $c$ with respect to $k_g$ and $k$ will not affect the
response. On the contrary, a decrease of $S$ and $c$ with respect
to $k_g$ and $k$ can drive the system into a non--saturated regime,
inhibiting the response mechanism.

\section{Results with no delay}
In the case that the production of mdm2 can be regarded as
instantaneous (no delay, $\tau=0$),
the concentration of p53 is rather insensitive to the change of the
dissociation constant $k$.
The stationary values of $p$ and $m$ are found as fixed points
of the equations \ref{eq1} (see Appendix) and
in Table I we list the stationary values $p^*$ of the amount of p53
molecules for
values of $k$ spanning seven orders of magnitude around the basal value
$k=180$.
Moreover, transient oscillatory behaviour upon change in the
dissociation
constant $k$ is not observed. This is supported by the fact that
the eigenvalues of the stationary points (listed in Table I) have
negative real parts, indicating stable fixed points, and rather
small imaginary pats indicating absence of oscillations.

More precisely, the variation $\Delta p$ of the stationary amount of
p53 if the
dissociation constant undergoes a change $\Delta k$ can be estimated,
under
the approximation that $k_g\ll p$  (cf. the Appendix), to be
\begin{equation}
\label{eq2}
\Delta p = \frac{d(S-bp)}{ack_g(a+b)}\Delta k.
\end{equation}
The fact that $\Delta p$ is approximately linear with $\Delta k$ with a
proportionality
constant which is at most of the order of $10^{-2}$  makes this system
rather
inefficient as response mechanism. Furthermore, it does not agree with
the
experimental data which show a peak of p53 followed, after several
minutes, by a peak
in mdm2 \cite{haupt}, and not just a shift of the two concentration to
higher values.

To check whether the choice of the system parameters affects the
observed behaviour,
we have repeated all the calculations varying each parameter of five
orders of magnitude
around the values used above. The results (listed in Table II for $S$
and $k_g$ and
not shown for the other parameters) indicate the same behaviour as
above (negative real part and no or small imaginary part in the
Eigenvalues). 
Consequently, the above results about the dynamics of p53 seem not to
be 
sensitive to the detailed choice of parameters (on the contrary, the
amount of mdm2
is quite sensitive). 

\section{Results with delay}
The dynamics changes qualitatively if we introduce a nonzero delay in
Eqs. \ref{eq1}.
Keeping that the halflife of an RNA molecule is of the order of 1200 s
\cite{holstege}, we repeat the calculations with $\tau=1200$. The Eqs. \ref{eq1} are
solved numerically, starting from the conditions $p(0)=0$ and $m(0)=0$
and making use of a variable--step Adams algorithm. 
After the system has reached its stationary state
under basal condition, a stress is introduced (at time $t=20000$ s) by
changing instantaneously the dissociation constant $k$. In Fig.
\ref{fig2}
we display a case in which the stress multiplies $k$ by a factor $15$
(a),
a case in which it divides it by a factor $15$ (b) and by a factor $5$
(c).

When $k$ is increased by any factor, the response is very similar to
the
response of the system without delay (cf. e.g. Fig. \ref{fig2}a). On
the
contrary, when $k$ is decreased the system displays an oscillatory
behaviour.
The height $\Delta p$ of the response peak is plotted in Fig.
\ref{fig3} as
a function of the quantity which multiplies $k$. If the multiplier is
larger
than $0.1$ the response is weak or absent. At the value $0.1$ the
system has a
marked response (cf. also Figs. \ref{fig2}b and c). The maximum of the
first
peak takes place approximately $1200$s after the stress, which is
consistent
with the lag--time observed in the experiment \cite{haupt},
and the peaks are separated from $\approx 2300$s.

Although it has been suggested that the effect of the stress is to
increase the
dissociation constant between p53 and mdm2 \cite{gottlieb}, our results
indicate that an efficient
response take place if $k$ decreases of a factor $\geq 15$ (cf. Fig.
\ref{fig2}b).
One has to notice that the conclusions of ref. \cite{gottlieb} have been
reached
from the analysis {\it in vivo} of the overall change in the
concentration of p53,
not from the direct measurement of the binding constant after
phosphorylation. Our
results also agree with the finding that p53asp20 (a mutated form of
p53 which mimicks 
phosphorylated p53, due to the negative charge owned by aspartic acid)
binds mdm2
{\it in vitro} more tightly than p53ala20 (which mimicks
unphosphorylated p53) \cite{gottlieb}.

This hypothesis is supported by molecular energy calculations made with
classical force fields.
Even if this kind of force fields is not really reliable for the
calculation of binding
constants, it gives
an estimate of the sign of the change in interaction among p53 and mdm2
upon phosphorylation.
We have performed an energy minimization of the conformation of the
system
composed by the binding sites of p53 and mdm2, starting form the
crystallographic
positions of ref. \cite{kussie} and using the force fields mm3
\cite{allinger} and mmff \cite{halgren},
for both the wild--type system and for the system where serine 20 of
p53 in phosphorylated.
Using mm3 we found that the phosphorylated system has an electrostatic
energy $16$ kcal/mol
lower than the wild--type system, while this difference is $26$
kcal/mol using the mmff
force field. Our calculations suggest that phopshprylated p53 is more
attracted by mdm2
due to the enhanced interaction of phosphorylated SER20 with LYS60,
LYS46 and LYS70 of mdm2,
and consequently the dissociation constant is lowered.

The robustness of the response mechanism with respect to the parameters
of the system, which is typical of many biological systems (cf., e.g.,
\cite{sou,leibler}), has 
been checked both to assess the validity of the model and to search for
weak points
which could be responsible for the upraise of the disease. Each
parameter has been varied
of five orders of magnitude around its basal quantity. The results are
listed in Table III.
One can notice that the response mechanism is quite robust to changes
in the parameters
$a$, $b$ and $c$. For low values of $a$ or $c$ the system no longer
oscillates, but displays,
in any case, a rapid increase in the amount of $p$ which can kill the
cell. This is true
also for large values of $d$. What is dangerous for the cell is a
decrease of $d$ or of $k_g$,
which would drop the amount of p53 and let the damaged cell survive.
This corresponds either
to an increase of the affinity between p53 and the mdm2 gene, or to an
increase of mdm2 half--life. 

To be noted that, unlike the case $\tau=0$, the system with delay never displays damped oscillations as a consequence of the variation of the parameters in the range studied in the present work. This sharp behaviour further testifies to the robustness of the response mechanism. Anyway, one has to keep in mind that the oscillating response produces the death of the cell, and consequently the long--time behaviour is only of theoretical interest.

The minimum value of the delay which gives rise to the oscillatory behaviour is $\tau\approx 100$s. For values of the delay larger  than this threshold, the amplitude of the response is linear with $\tau$ (cf. Fig. 4), a fact which is compatible with the explanation of the response mechanism of Section IV.

The lag time before the p53 response is around $3000$s (in accordance
with the 1h delay
observed experimentally \cite{haupt} and is independent on all
parameters, except $c$ and $\tau$. The dependence of the lag time
on $\tau$ is approximately linear up to $5000$s (the longest delay
analyzed).
Increasing $c$ the lag time increases to $8000$s (for $c=10^4$)
and $25000$s (for $c=10^5$). On the other hand, the period of
oscillation depends only on
the delay $\tau$, being approximately linear with it.

We have repeated the calculations squaring the variable $p$ in Eqs. 1, to keep into account the cooperativity induced by the fact that the active form of p53 is a dimer of dimers \cite{fersht}. The results display qualitative differences neither for non--delayed nor for the delayed system.

\section{Discussion}

All these facts can be rationalized by analyzing the mechanism which
produces the response.
The possibility to trigger a {\sl rise} in p53 as a dynamic response
to an {\sl increased} binding between p53 and mdm2,
relies on the fact that a sudden increase in p--m binding diminishes the
production of $mdm2$,
and therefore (subsequently) diminishes the amount of $pm$.
In other words, while the change in $k$ has no direct effect in the 
first of Eqs. \ref{eq1}, it directly reduces mdm2 production by
subtracting p53 from the gene which producese mdm2. 

Mathematically, the oscillations arise because the saturated nature of
the 
binding $pm$ imply that
pm is approximately equal to the minimum between p and m. Each time the
curves associated with p and m
cross each other (either at a given time or $\tau$ instants before),
the system has 
to follow a different set of dynamic equations than before,
finding itself in a state far from stationarity. This gives rise to the
observed peaks.

To be precise, the starting condition (before the stress) is $m>p$.
The stress reduces the dissociation constant $k$ of, at least, one
order of magnitude, causing
a drop in $p$, which falls below $m$. For
small values of $k$ (to be precise, for $k\ll \min(|p-m|,p,m)$), one
can make
the simplification $pm\approx \min(p,m)$, and consequently rewrite Eqs.
\ref{eq1} as
\begin{eqnarray}
\label{simply1}
\mbox{for $p<m$}&&\frac{\partial p}{\partial t} =S-(a+b)p\\
\label{simply2}
\mbox{for $p(t-\tau)<m(t-\tau)$}&&\frac{\partial m}{\partial t}=-dm \\
\label{simply3}
\mbox{for $p>m$}&&\frac{\partial p}{\partial t}=S-am-bp\\
\label{simply4}
\mbox{for $p(t-\tau)>m(t-\tau)$}&&\frac{\partial m}{\partial
t}=c\frac{p(t-\tau)-m(t-\tau)}{k_g+p(t-\tau)-m(t-\tau)}-dm.
\end{eqnarray}
Each period after the stress can be divided in four phases. In the
first one $p<m$ and $p(t-\tau)<m(t-\tau)$, 
so that $p$ stays constant at its stationary value $S/(a+b)\approx
p^*$, while $m$ decreases with
time constant $d^{-1}$ towards zero (not exactly zero, since the
approximated Equations do not
hold for $m\sim k$). 
In the second phase one has to consider the second 
($p(t-\tau)<m(t-\tau)$) and the third ($p>m$) Equations (\ref{simply2}
and \ref{simply3}). 
The new stationary value for $p$ is $(S-am)/b\approx S/b$ which
is much larger than $p^*$ since $b\ll a$. This boost takes place in a
time of the
order of $b^{-1}$, so if $b^{-1}>\tau$, as in the present case, $p$ has
no time to reach the
stationary state and ends in a lower value. In the meanwhile, $m$
remains in the low value given by
Eq. \ref{simply2}. At a time $\tau$ after the stress Eq. \ref{simply2}
gives way to
Eq. \ref{simply4}. 
The latter is composed by a positive term which is $\approx c$ if 
$p(t-\tau)-m(t-\tau)\gg k_g$ and $\approx 0$ under the opposite
condition. 
Since $p(t-\tau)\gg m(t-\tau)$ (it refers to the boost of $p$), 
then the new stationary value of $m$ is $c/d\approx m^*$. 
The raise of $m$ takes place in a time of the
order of $d^{-1}$ and causes the decrease of $p$, whose production rate
is ruled by $S-am$. 
The fourth phase begins when $p$ approaches $m$. 
Now one has to keep Eqs. \ref{simply1} and \ref{simply4}, 
so that $p$ returns to the basal value $p^*$, while $m$ stays for a
period of 
$\tau$ at the value $c/d\approx m^*$ reached 
in the third phase. After such period, Eq. \ref{simply2} substitutes
Eq. \ref{simply4} and another
peak takes place.

The heigth of the p53 peak is given by $S/b$ if $p$ has time to reach
its stationary state
of phase two (i.e., if $b^{-1}<\tau$), or by $S/b(1-\exp(-b\tau))$ if
the passage to the
third phase takes place before it can reach the stationary state. The
width of the peak is
$\approx \tau$ and the spacing among the peaks $\approx \tau$, so that
the oscillation period
is $\approx 2\tau$.

The necessary conditions for the response mechanism to be effective are
1) that $s/a\ll c/d$, that is
that the stationary value of $p$ just after the stress is much 
lower than the stationary value of $m$, 2)
that $b\ll a$, in such a way that the stationary state of $p$ in the
second phase is much larger
than that in the first phase, in order to display the boost, 3) that
$d^{-1}<\tau$, otherwise
$m$ has not enough time to decrease in phase one and to increase in
phase three. 

The failure of the response for low values of $a$ (cf. Table 3) 
is due to the fall of condition 2), the failure for small $c$
is caused by condition 1), the failure at small and large values of $d$
is associated with conditions 3)
and 1), respectively. At low values of $k_g$ the response does not take
place because the
positive term in Eq. \ref{simply4} is always $\sim c$, and thus $m$
never decreases.

\section{Conclusions}
In summa, we have shown that the delay is an essential ingredient of
the system to have a ready and robust
peak in p53 concentration as response to a damage stress. In order to
have a peak which
is comparable with those observed experimentally, the dissociation
constant between p53 and mdm2
has to decrease of a factor $15$. Although it is widely
believed that phosphorylation of p53 increases the dissociation
constant, we observe an oscillating
behaviour similar to the experimental one only if $k$ decreases. In this case the
response
is quite robust with respect to the parameters, except upon increaasing
of the half--life
of mdm2 and upon decreasing of the dissociation constant between p53
and the mdm2 gene, in which
cases there is no response to the stress. Moreover, an increase in the
production rate of
mdm2 can delay the response and this can be dangerous to the cell as
well.
We hope that detailed experimental measurements of the physical
parameters of the system will be made soon,
in order to improve the model and to be able to make more precise
predictions about the weak
point of the mechanism, weak points which could be intimately connected
with the upraise of cancer.

\newpage
\centerline{{\Large Appendix}}
\bigskip

The stationary condition for Eqs. \ref{eq1} without delay can be found
by the intersection of the curves
\begin{eqnarray}
m(p)&=&\frac{c(a+b)p-cs}{d(a+b)p-d(S-a k_g}\\
m_k(p)&=&\frac{(S-bp)((a+b)p+ak-S)}{a((a+b)p-S)},
\end{eqnarray} 
which have been obtained by the conditions ${\dot p}={\dot m}=0$, 
explicitating $pm$ from the first of Eqs. \ref{eq1} and
substituting it in the second and the third, respectively.
To be noted that $m_k$ is linear in $k$.

The variation $\Delta p$ of the stationary value of p53 upon change
$\Delta k$ in the
dissociation constant can be found keeping that
\begin{equation}
\frac{dp}{dk}=\frac{dp}{dm}\frac{dm_k}{dk}\approx
\frac{d(S-bp)}{ck_g(a+b)},
\end{equation}
where the approximation $k_g\ll p$ has been used.
Consequently,
\begin{equation}
\Delta p=\frac{d(S-bp)}{ck_g(a+b)}\Delta k,
\end{equation}
which assumes the largest value when $p$ is smallest. Using the
parameters listed above, the
proportionality constant is, at most, $10^{-2}$.

Furthermore, keeping that $pm<\min(p,m)$ for any value of $p$ and $m$,
the Eigenvalues of
the dynamical matrix have negative real part, indicating that the
stationary
points are always stable.

\newpage

\begin{table}
\caption{Stationary values $p^*$ and $m^*$ for the amount of p53 and
mdm2, respectively,
calculated at $\tau=0$. In the last column the Eigenvalues of the
linearized (around the fixed points $p^*$, $m^*$) dynamical
matrix are displayed, by real and imaginary part. 
The real part of the Eigenvalues is always negative
and the imaginary part, when different from zero, is lower than the real
part, indicating that the stationary values are always stable and the
dynamics is overdamped.}
\begin{tabular}{|c|c|c|c|}
\hline
$k$ & $p^*$ & $m^*$ & $\lambda_{1,2}$ \\\hline
0.18 & 47.3 & 33.6 & -0.017$\pm$ 0.013i \\
1.8 & 49.5 & 36.8 & -0.012,\, -0.014 \\
18 & 63.9 & 52.4 & -0.011,\, -0.008 \\
{\bf 180} & {\bf 154.6} & {\bf 81.3} & {\bf -0.007$\pm$0.001i} \\
1800 & 858.3 & 96.7 & -0.009,\, -0.0008 \\
18000 & 4287 & 99.3 & -0.009,\, -0.0002 \\
180000 & 8632 & 99.6 & -0.009,\, -0.0001 \\\hline
\end{tabular}
\end{table}

\begin{table}
\caption{Same as in Table I, varying some of the parameters which
define the system of
five orders of magnitude. In each cell it is indicated the quantity at
$k=18$, $k=180$
(basal value) and $k=1800$. }
\begin{tabular}{|c|c|c|c|}
\hline
  & $p^*$ & $m^*$ & $\lambda_{1,2}$ \\\hline
$S=0.01$ & 1.6 & 4.6 & $-0.02\pm 0.006i$ \\
       & 4.6 & 13.6 & -0.007,\, -0.005\\
       & 15.1 & 34.6 & -0.009,\, -0.001 \\
$S=0.1$ & 8.3 & 15.1 & $-0.01\pm 0.008i$ \\
        & 20.2 & 37.8 & $-0.007\pm 0.001i$ \\
        & 81.1 & 73.6 & -0.009,\, -0.001 \\ 
$S=1$ & 63.9 & 52.4 & $-0.01\pm 0.008i$ \\
       & 154.6 & 81.3 & $-0.007\pm 0.001i$ \\
       & 858 & 96.7 & -0.009,\, -0.0008 \\
$S=10$ & $70019$ & $99.9$ & -0.009,\,-$10^{-4}$ \\
         & 70088 & $99.9$ & -0.009,\,-$10^{-4}$\\
         & 70756 & $99.9$ & -0.009,\,-$10^{-4}$\\
$S=100$ & 970000 & $99.9$ & -0.01,\,-$10^{-4}$ \\
         & 970000 & $99.9$ & -0.01,\,-$10^{-4}$ \\
          & 970000 & $99.9$ & -0.01,\,-$10^{-4}$ \\
\hline
$k_g=0.28$ & 42.5  & 97.0 & -0.02,\, -0.01 \\
           & 121.7 & 99.6 & -0.009,\, -0.006 \\
           & 824.1 & 99.9 & -0.009,\, -0.006 \\
$k_g=2.8$ & 45.5 & 81.5 & $-0.01\pm 0.003i$ \\
          & 125.2 & 97.0 & -0.009,\, -0.006 \\
          & 827.3 & 99.6 & -0.009,\, -0.008 \\
$k_g=28$ & 63.9 & 52.4 & $-0.01\pm 0.008i$ \\
         & 154.6 & 81.3 & $-0.007\pm 0.001i$  \\
         & 858 & 96.7 & -0.009,\, -0.0008  \\
$k_g=280$ & 192.7 & 36.3 & $-0.006\pm 0.0003i$ \\
          & 331.1 & 51.6 &  -0.008,\, -0.003  \\
          & 1104.1 & 79.3 & -0.009,\, -0.007  \\
$k_g=2800$ & 1214 & 29.0 & -0.009,\,  -0.0006 \\
           & 1380 & 32.5 & -0.009,\, -0.0006  \\
           & 2345 & 45.3 & -0.009,\, -0.0006  \\
\hline
\end{tabular}
\end{table}

\begin{table}
\caption{The value of $\Delta p/p$ when the parameters a, b, c, d and
$k_g$ are scaled
of the quantity listed in the first column. (1) indicates that the
system does
not oscillate and $p$ reaches a stationary value much larger than
before the stress.
(2) indicates that the system does not display any response to the
stress or displays a
negative response. The star
indicates that the peak appears after 4000s (for $c=10^4$) and 25000s
(for $c=10^5$).}
\begin{tabular}{|c|c|c|c|c|c|}
\hline
scale & $a$ & $b$ & $c$ & $d$ & $k_g$ \\\hline
$10^{-2}$ & (1) & $3.43$ & (1) & (2) & (2) \\
$10^{-1}$ & (1) & $3.46$ & (1) & (2) & (2) \\
1 & $3.2$ & $3.2$ & 3.2 & 3.2 & 3.2 \\
$10$ & $11.1$ & $2.14$ & 9.2* & (1) & 3.2\\
$100$ & $2.1$ & (2) & 1.4* & (1) & (2) \\\hline
\end{tabular}
\end{table}

\newpage
\begin{figure}
\centerline{\psfig{file=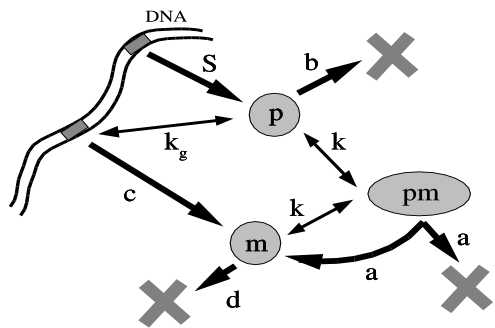,height=11cm,width=9cm}}
\caption{A sketch of the loopback mechanism which control the amount of
p53 in the cell. The grey crosses indicate that the associated molecule
leaves the system.}
\label{fig1}
\end{figure}

\begin{figure}
\centerline{\psfig{file=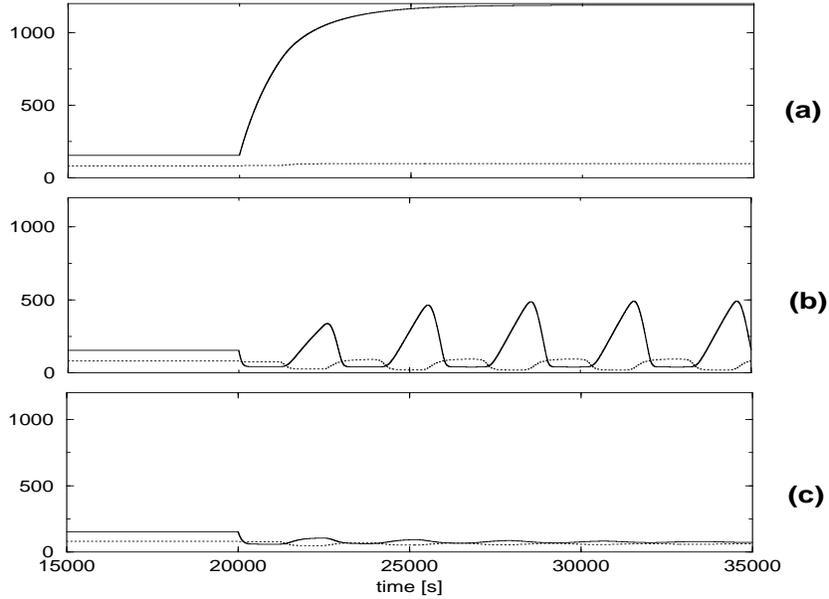,height=8cm,width=11cm,angle=-90}}
\caption{The response in the concentration of p53 (solid line) and mdm2
(dotted line)
upon variation of the dissociation constant $k$. At time 20000 s the
constant $k$
is multiplied by 15 (a), divided by 15 (b) and divided by 5 (c).}
\label{fig2}
\end{figure}

\begin{figure}
\centerline{\psfig{file=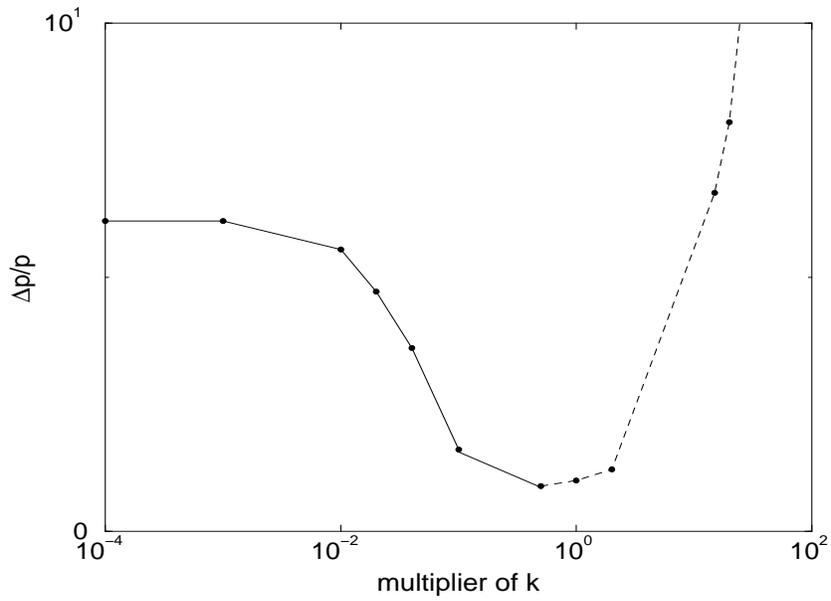,height=8cm,width=11cm,angle=-90}}
\caption{The height of the response peak $\Delta p$ with respect to the
quantity that multiplies $k$, mimicking the stress. The dotted line
indicates that the system does not display oscillatory behaviour.}
\label{fig3}
\end{figure}

\begin{figure}
\centerline{\psfig{file=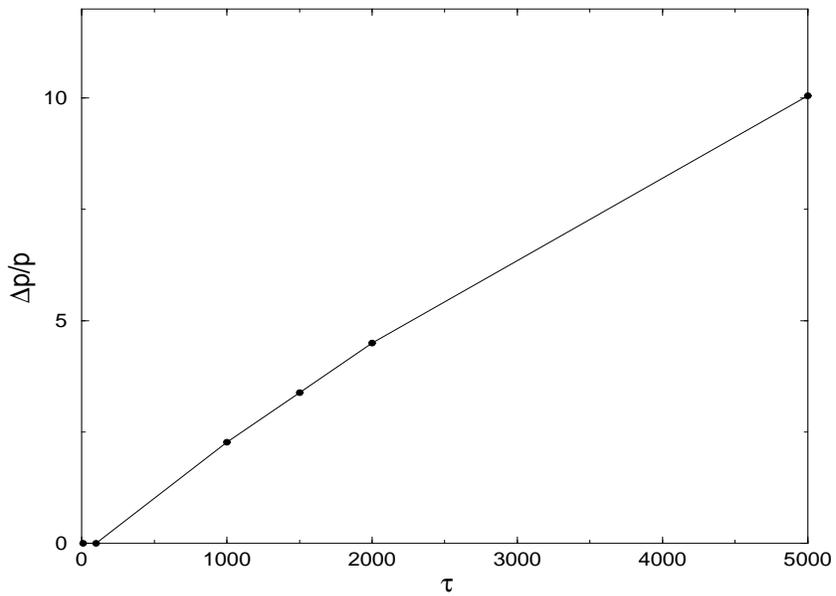,height=8cm,width=11cm,angle=-90}}
\caption{The dependence of the  height of the response peak $\Delta p$
on the delay $\tau$.}
\label{fig4}
\end{figure}

\end{document}